\documentclass[prc,aps,twocolumn,showpacs,preprintnumbers,eqsecnum,amsmath,amssymb,twocolomn]{revtex4-1}
\usepackage{graphicx}
\usepackage{epsfig}
\usepackage{dcolumn}
\usepackage{bm}
\begin{document}
\title{New Modes of Nuclear Excitations}

\author{
 N.~Tsoneva$^{1,2}$, H.~Lenske$^{1}$}
\affiliation{
  $^1$Institut f\"ur Theoretische Physik, Universit\"at Gie\ss en,
  Heinrich-Buff-Ring 16, D-35392 Gie\ss en, Germany \\
$^2${Institute for Nuclear Research and Nuclear Energy, 1784 Sofia, Bulgaria}}

\begin{abstract}
We present a theoretical approach based on density functional theory supplemented by a microscopic multi-phonon model which is applied for investigations of pygmy resonances and other excitations of different multipolarities in stable and exotic nuclei. The possible relation of low-energy modes to the properties of neutron or proton skins is systematically studied in isotonic and isotopic chains. 
The fine structure of nuclear electric and magnetic response functions is analyzed and compared to experimental data. Their relevance to nuclear astrophysics is discussed.
\end{abstract}
\maketitle

\section{Introduction}
\label{intro}
 The precise knowledge of nuclear response functions plays a key role in the determination of photonuclear reactions cross sections which are of importance for the nucleosynthesis of heavier elements. In this connection information on low-energy excitations located  around the neutron threshold is needed. 
During the last decade, a new low-energy mode called pygmy dipole resonance (PDR) has been observed \cite{Sav13}. It was intensively investigated, in many experimental and theoretical studies where unknown aspects on the isospin dynamics of the nucleus have been revealed \cite{Sav13}. As the most unique feature of the PDR was pointed out its close connection to nuclear skin oscillations which become visible in transition densities and currents \cite{Rye02,Paa05,Tso08}.  

Recently, in theoretical studies a higher order multipole pygmy resonance, namely the pygmy quadrupole resonance (PQR) has been predicted \cite{Tso11}.
Here, our theoretical investigations on dipole and other multipole excitations in stable and exotic nuclei, particularly exploring their connection to neutron/proton binding energies and thickness of the neutron (proton) skin are discussed.
In these studies, our theoretical method based on self-consistent Hartree-Fock-Bogoljubov (HFB) description of the nuclear ground state \cite{Hof98} and quasiparticle-phonon model (QPM) \cite{Sol76} for the excited states is applied \cite{Tso11,Tso08,Tso04,Vol06,Schw13}. As a link to nuclear many-body theory a density functional theory (DFT) is used \cite{HoKohn:64}. The approach gives us a full flexibility to describe the nuclear ground state properties like binding energies, neutron and proton root mean square radii and the difference between them defining the nuclear skin, and separation energies to the required accuracy in medium and heavy nuclei \cite{Tso08,Tso11}.
Of special importance for the present investigations are the nuclear surface regions, where the formation of a skin takes place \cite{Vol06,Tso08,Schw13}.

\section{The Theoretical Model}

\label{sec-1}
An important advantage is the description of the nuclear excitations in terms of quasiparticle-random-phase- approximation (QRPA) phonons as a building blocks of the QPM model space \cite{Sol76} which provides a microscopic way to multi-configuration mixing. The QPM allows for sufficiently large configuration spaces which is required for a unified description of low-energy single- and multi-particle states and also for the GDR energy regions.

In this connection, for spherical even-even nuclei the model Hamiltonian is diagonalized on an orthonormal set of wave functions constructed from one-, two- and three-phonon configurations \cite{Gri94}.

\begin{equation}
\Psi_{\nu} (JM) =
 \left\{ \sum_i R_i(J\nu) Q^{+}_{JMi}
\right.
+ \sum_{\scriptstyle \lambda_1 i_1 \atop \scriptstyle \lambda_2 i_2}
P_{\lambda_2 i_2}^{\lambda_1 i_1}(J \nu)
\left[ Q^{+}_{\lambda_1 \mu_1 i_1} \times Q^{+}_{\lambda_2 \mu_2 i_2}
\right]_{JM}
\label{wf}
\end{equation}
\[
\left.
{+ \sum_{_{ \lambda_1 i_1 \lambda_2 i_2 \atop
 \lambda_3 i_3 I}}}
{T_{\lambda_3 i_3}^{\lambda_1 i_1 \lambda_2 i_2I}(J\nu )
\left[ \left[ Q^{+}_{\lambda_1 \mu_1 i_1} \otimes Q^{+}_{\lambda_2 \mu_2
i_2} \right]_{IK}
\otimes Q^{+}_{\lambda_3 \mu_3 i_3}\right]}_{JM}\right\}\Psi_0
\]
where R, P and T are unknown amplitudes, and $\nu$ labels the
number of the  excited states.

The electromagnetic transition matrix elements are calculated for transition operators including the interaction of quasiparticles and phonons \cite{Pon98} where exact commutation relations are implemented which is a necessary condition in order to satisfy the Pauli principle.

\section{Discussion}
\label{sec-2}
\subsection{First Systematic Studies of Pygmy Dipole Resonance in N=50 Isotones}

\begin{figure*}
\begin{center}
\resizebox{2.0\columnwidth}{!}{
\includegraphics[width=6.5cm,clip]{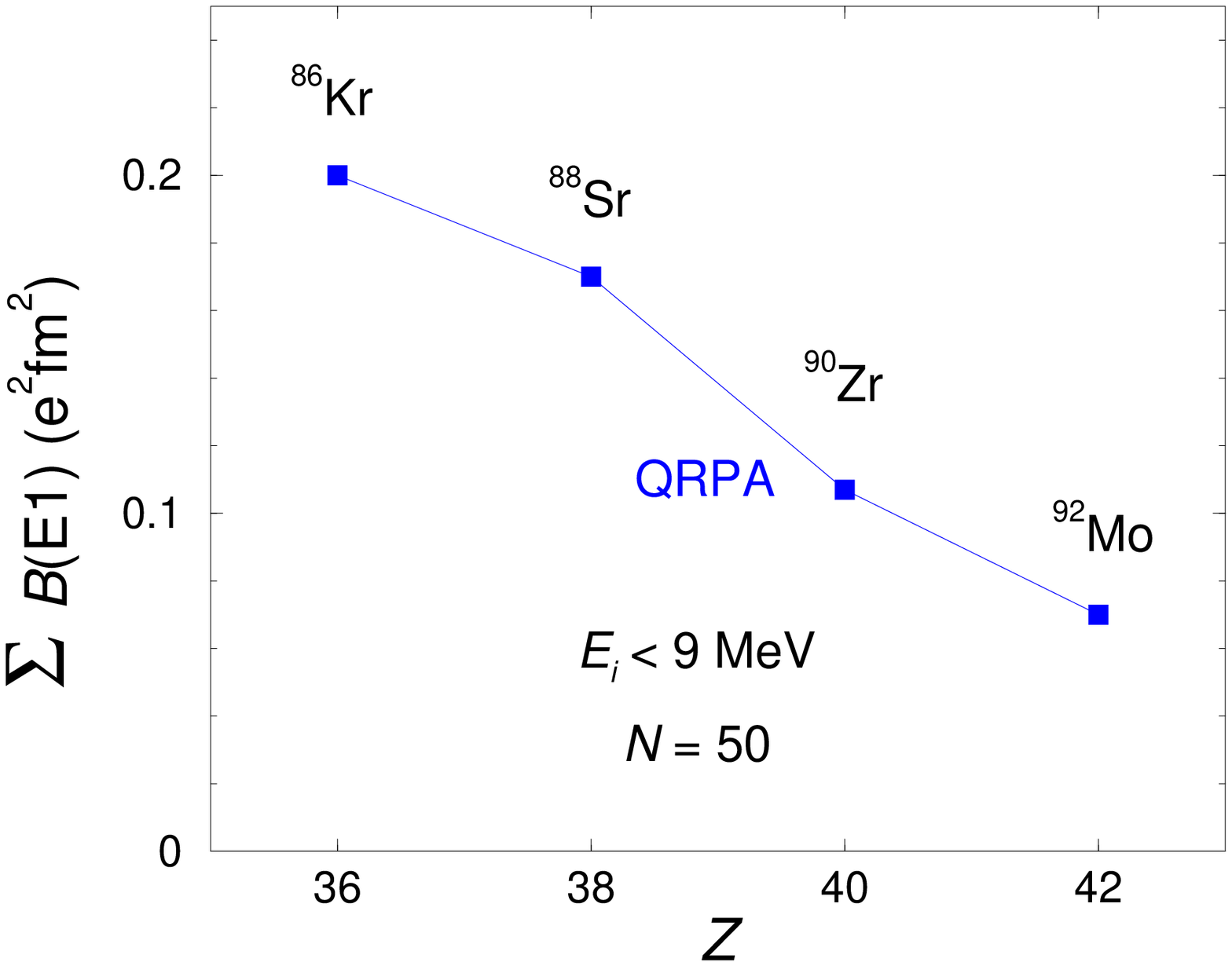}
\includegraphics[width=6.5cm,clip]{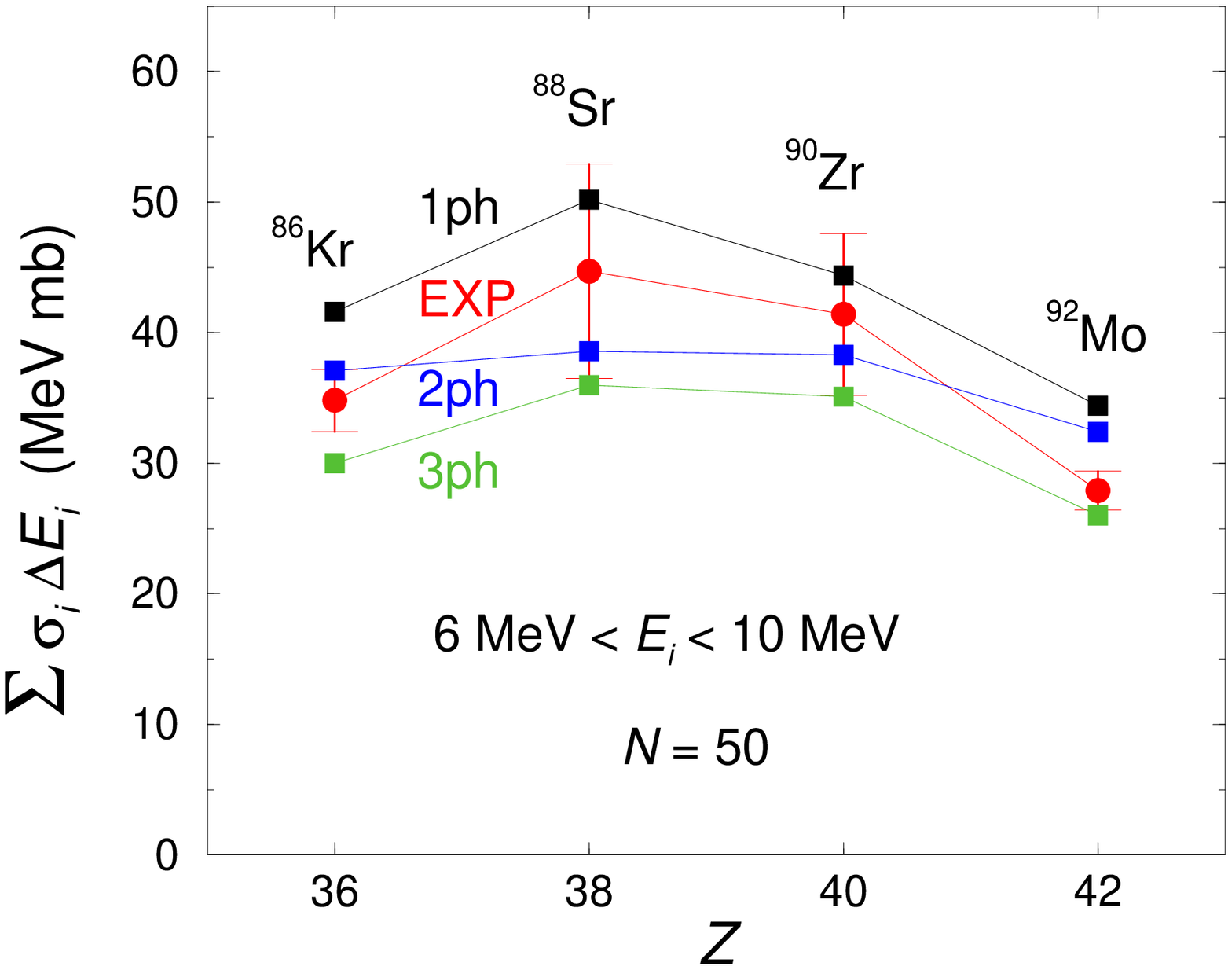}}
\caption{(Color online) 
(left panel) Total B(E1) strength up to 9 MeV obtained from QRPA calculations for N=50 isotones [8].
(right panel) Energy-weighted sums of photoabsorption cross sections for the excitation energy region from 6 to 10 MeV
for N=50 isotones [8].}
\label{fig-1}       
\end{center}
\end{figure*}

First systematic experimental and theoretical studies of the electric dipole response of $N=50$ isotones - $^{88}$Sr, $^{90}$Zr, $^{92}$Mo and $^{86}$Kr up to 10 MeV are done \cite{Schw13} (see in Fig. \ref{fig-1} (right panel)).
Experimental data and QPM calculations reveal a strong enhancement of the  E1 strength in the energy range E*=6-10  MeV with respect to  the shape of a  Lorentz-like strength function used to adjust the GDR \cite{Schw13}.
From QRPA calculations shown in Fig. \ref{fig-1} (left panel), the energy region below E*$\leq$9 MeV is related to PDR \cite{Tso08,Schw13} whose total strength smoothly decreases with increasing Z number closely correlated with the thickness of the neutron skin \cite{Schw13}.
For $^{86}$Kr nucleus the experiment identifies 42 levels up to an excitation energy of 10.1 MeV. Microscopic three-phonon QPM calculations of the electric dipole strength function of this nucleus successfully describe the experimental data \cite{Schw13}.  

\begin{figure*}
\begin{center}
\resizebox{2.0\columnwidth}{!}{
\includegraphics[width=6.5cm,clip]{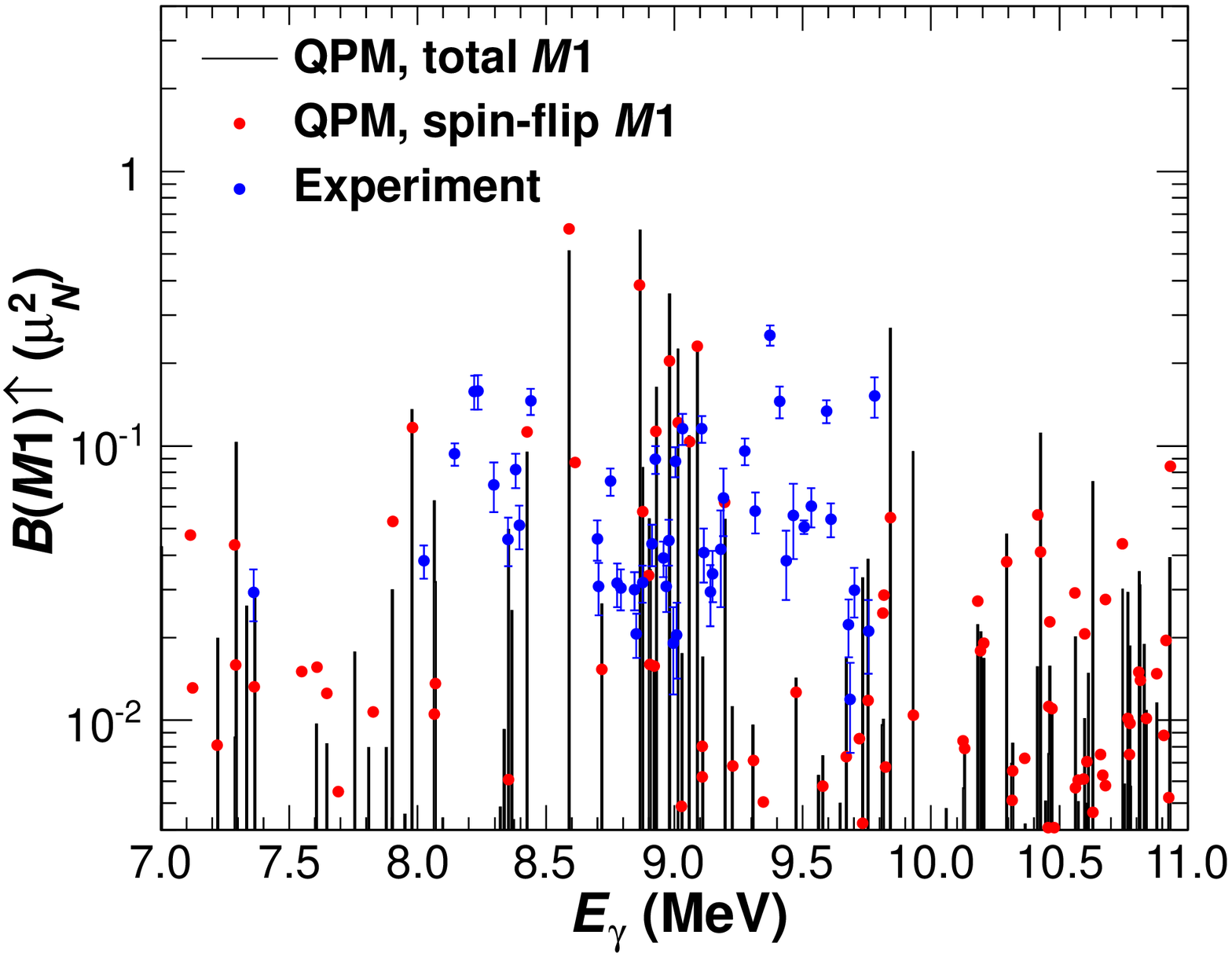}
\includegraphics[width=6.5cm,clip]{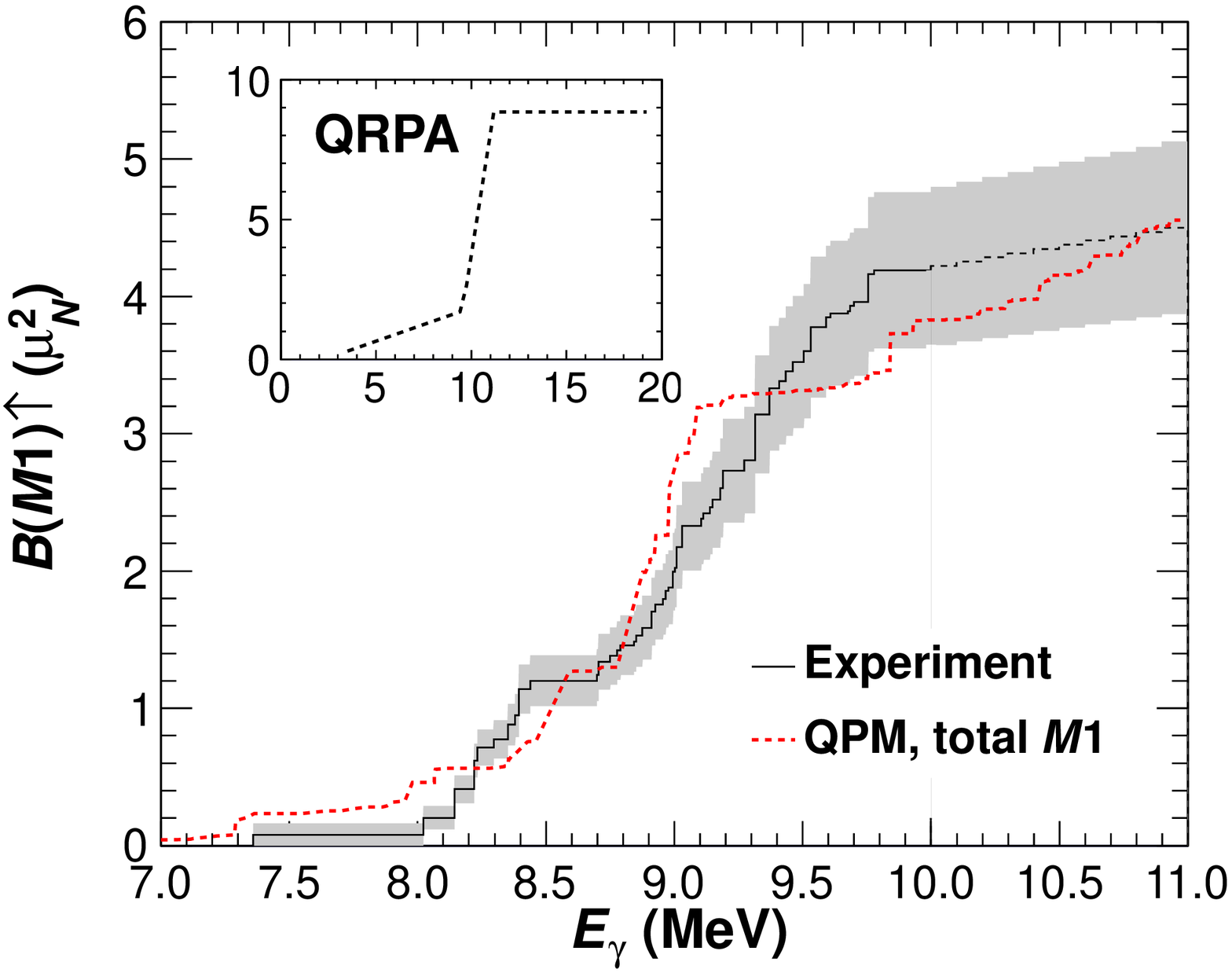}}
\caption{(color online). (left panel) The measured strength of discrete levels in $^{90}$Zr
compared predictions from the quasiparticle-phonon model;
(right panel) A comparison of the measured and calculated QPM cumulative M1 strength. The dashed line
continuing the solid line above 10 MeV represents the M1 strength obtained from the continuum.The shaded area gives the uncertainty
of the experimental values. QRPA results of the total cumulative M1 strength up to 20 MeV are shown in the insert}
\label{fig-2}  
\end{center}     
\end{figure*}

The present analysis also shows that theoretical standard strength functions currently used for the calculation of cross sections in codes based on statistical reaction models do not describe the dipole strength distribution below the $(\gamma,n)$ threshold correctly and need to be improved by taking into account the observed enhanced strength. In this aspect, the three-phonon QPM
could be successfully implemented in statistical reaction codes to investigate neutron-capture cross sections of astrophysical importance  \cite{Raut13}.

\subsection{Distinguishing of the Pygmy Dipole Resonance from other Low-Energy Excitations}
A major experimental problem is to distinguish
$M$1 and $E$1 strength, since both of them are highly fragmented at these energies.
In order to unambiguously discriminate between these dipole excitation modes,
the spin and parity of the individual states must be known. That was achieved in high-sensitivity studies of E1 and M1 transitions observed in the $^{138}$Ba($\vec{\gamma}$,$\gamma$') reaction at energies below the neutron emission threshold which have been performed using the nearly monoenergetic and 100\% linearly polarized photon beams from the HI$\vec{\gamma}$S facility \cite{Pie01,Ton10}.
The electric dipole character of the so-called $pygmy$ mode was experimentally verified for excitations from 4.0 - 8.6 MeV \cite{Ton10}.
The fine structure of the $M$1 $spin-flip$ mode was observed for the first time in $N$ = 82 nuclei.
The data could be very well explained by our QPM calculations of low-energy E1 and M1 strengths in $^{138}$Ba \cite{Ton10}.

Overall, our theoretical results are in good agreement with the present E1 and M1 data below the particle threshold, both with respect to the centroid energies and summed transition strengths. A common feature of the low-energy $1^-$ and $1^+$ states is that both modes are excited by almost pure two-quasiparticle (2-QP) QRPA states. They serve as doorway states which then decay into multi-configuration states with complicated multi-phonon wave functions, thus giving rise to fragmentation of the spectral distributions \cite{Ton10}.
As the excitation energy is increased, the isovector contribution to the dipole strength increases following closely its Lorentzian fall-off often assumed with GDR in data analyses \cite{Ber75}. Theoretically, this can be seen in transition densities and state vectors structure which manifest an enlarging of the out-of-face neutron to proton contributions and corresponding energy-weighted sum rules which is generally associated with the GDR \cite{Tso08,Schw13}.

\subsection{Description of the Fine Structure of the M1 Spin-Flip Resonance in  $^{90}$Zr }

Recently, the first high-resolution photon-scattering experiment with monoenergetic and linearly polarized beams from 7 to 11 MeV has been performed in order to study the fine structure of the $M1$-Giant Resonance (GR) in the nuclide $^{90}$Zr \cite{Rus13}. The measurements reveal the fine structure of the giant $M1$ resonance with centroid energy of 9 MeV and sum strength of 4.5(4) $\mu^2_N$ which is further confirmed in our three-phonon QPM calculations and explained as fragmented spin-flip excitations \cite{Rus13}. The fragmentation pattern and absolute value of the $M1$ strength are explained. The theoretical investigations indicate a strong increase of the 
contribution of the orbital part of the magnetic moment due to coupling of 
multiphononon states which is an interesting result not reported before. The effect is estimated to account for about 22\% of the 
total $M1$ strength below the threshold. The good agreement of the calculated 
and measured total strengths is a signature that the quenching is handled 
reliably in the chosen approximation \cite{Rus13}. The results from the comparison between experiment and theory are presented in 
Fig. \ref{fig-2}.

\section{Conclusion}
The agreement between data and calculations confirms the predictive power of the QPM many-body theory for exploratory investigations of new modes of excitation \cite{Tso08,Tso11,Tso04,Vol06,Schw13,Ton10,Rus13}. The involvement of our approach in recent studies of n-capture reaction rates which are of astrophysical importance \cite{Raut13} could be further directed toward investigations of hitherto experimentally inaccessible mass regions.

\end{document}